\documentstyle[12pt]{article}
\input{psfig.sty}

\begin{document}

\title{A Review of The O--C Method and Period Change
\footnote{published in 1999, Publ. Beijing Astron. Obs., 33, p.17 }}
\author{ZHOU Ai-Ying\\
Beijing Astronomical Observatory, Chinese Academy of Sciences\\
E-mail: aiying@bao.ac.cn\\}
\date{}

\maketitle

\begin{abstract}
The classical O--C curves are discussed in different cases in which various
period changes involved. Among them, the analytic O--C curves with frequency,
amplitude modulations and with double modes are closely inspected, respectively.
As a special, the light-time effect is illustrated. The features of period
change noise and period change to metallicity are added at the end.
\end{abstract}

\bf
Keywords:~~\rm O--C method--period change\\

\section{Introduction}
\noindent The O--C diagram is a plot showing the observed times of
maximum light(O) minus those calculated according to an adopted
ephemeris(C) plotted as a function of time, mostly, the number of
elapsed cycles. In the same way, the O--C diagram can also be
constructed by the difference between the observed times of
maximum radial velocity and the times predicted from an adopted
ephemeris. One may find the minima are used instead of maxima for
some variables. In particular, the spectroscopic and photometric
O--C values are combined to produce a single O--C diagram.

The employment of the O--C diagram almost means normal or regular periodic
light curve with a large amplitude is concerned. In other words, the times
of maximum light can be determined sufficiently well from the observed light
data. One can find `O' by fitting a single sinusoid with an assumed pulsation
period to observations. `O' may be derived through local fit to the
light curves around individual maxima as well. The latter is very useful in
the case of asymmetrical light curves or light curves with changing amplitude
such as the cycle-to-cycle or the day-by-day variations.

The Calculated times of maximum light ($C_{n}$) based on an adopted or
estimated period ($P_{\rm est}$) and an initial epoch or Time of maximum
light ($T_{0}$). Assuming that the observed times of
maximum light, ($T_{0}, T_{1}, T_{2}, \cdots, T_{N}$), are determined in
different ways depending on different observers, we give O--C series as
\begin{equation}
  (O-C)_{n}= T_{n}-(T_{0}+ n\,P_{\rm est}), ~~~~n=0, 1, 2,\cdots\cdots, N
\label{eq:OCn1}
\end{equation}
where n is the cycle number elapsed from the initial epoch at the observed
point. Followings refer to different cases of period change in the point
of O--C curve.

\section{Different O--C Curves}
\subsection{Period Changing Nothing}
\begin{figure}
 \vspace{-25mm}
  \begin{center}
  \hspace{2mm}\psfig{figure=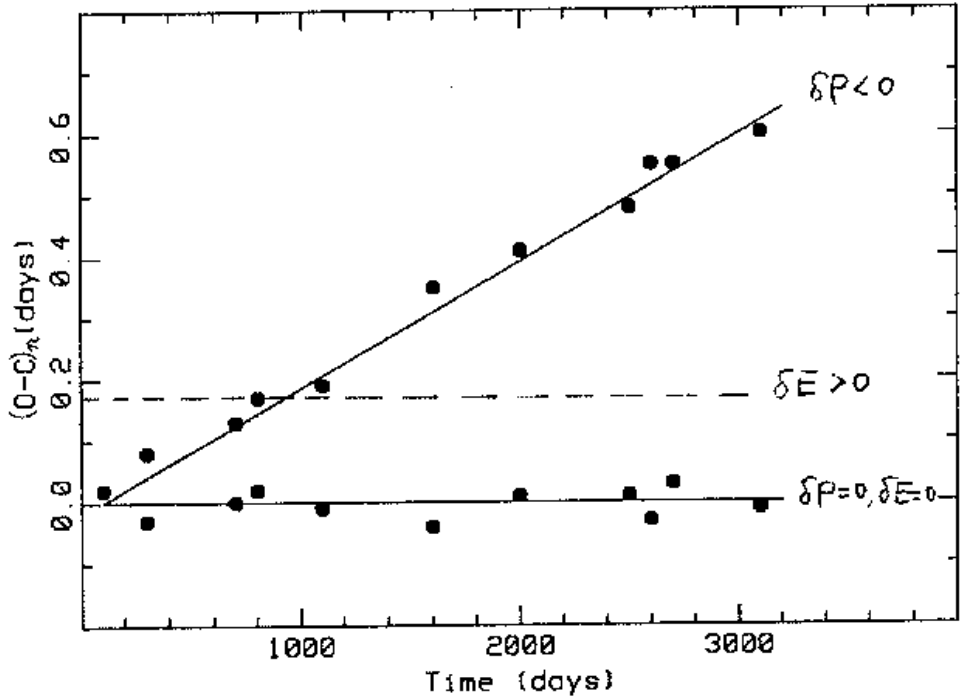,width=133mm,height=100mm,angle=-0.0}
  \end{center}\vspace{-5mm}
  \caption{The O--C diagram as a constant period. The slop line,
$(O-C)_{n}=n\,\delta P$, refer to the estimated period $P_{\rm est}=P+ \delta P$
assuming the intrinsic period P is constant. Here, $\delta$P=O(P)$\ll$ P.
Dashed lines, $(O-C)_{n}= -\delta T_0$, correspond to the cases in which
a correct period value is employed but the initial epoch has an
uncertainty of $\delta T_0$. Note that the mirror parts are omitted.}
\label{fig:OC1}
\end{figure}

\noindent
At the beginning, we consider the case of period P keeps constant over the
time span of observations covered. By assuming that
\begin{equation}
\begin{array}{l}
  P_{\rm est}=P + \delta P, ~~~T_{0}^{\rm est}=T_{0} + \delta T_{0}
\label{eq:PEest}
\end{array}
\end{equation}
then we can compute O--C as
\begin{equation}
  (O-C)_{n}= T_{n}-[T_{0}+\delta T_{0} +n(P+\delta P)]
\label{eq:OCn2}
\end{equation}
If $\delta$P=0, namely
\begin{equation}
\begin{array}{l}
  (O-C)_{n}=T_{n}-(T_{0}+\delta T_{0} + n P), ~~~\langle (O-C) \rangle = -\delta T_{0}
\end{array}
\end{equation}
the O--C diagram will consist of points scattered around a horizontal line:
\begin{equation}
  (O-C)_{n}= -\delta T_{0}                     \label{eq:OCn3}
\end{equation}
by assuming $T_{n}-(T_{0}+nP)=O(0.0)$, the same below.\\
If $\delta T_0$=0, that is
\begin{equation}
\begin{array}{l}
  (O-C)_{n}=T_{n}-(T_{0}+nP)-n\,\delta P          \label{eq:OCn4}
\end{array}
\end{equation}
Clearly, $(O-C){_n}$ decreases with n increases, the O--C plot will be a
straight line:
\begin{equation}
      (O-C)_{n}=-n\,\delta P                    \label{eq:OCn5}
\end{equation}
If $\delta P>0$, that is $P_{\rm est}$ is longer than P, the O--C line will slop
downwards; otherwise, $\delta P<0$, slop upwards. A vertical shift of the
$(O-C)_{n}$ line relative to $\delta$P=0, $\delta T_{0}$=0 lines in
Fig.~\ref{fig:OC1} tells us a shift in the epoch, the change in the
phase of variation in other words.
%
%

\subsection{Period Changing with Time}
\begin{figure}
  \begin{center}
  \hspace{2mm}\psfig{figure=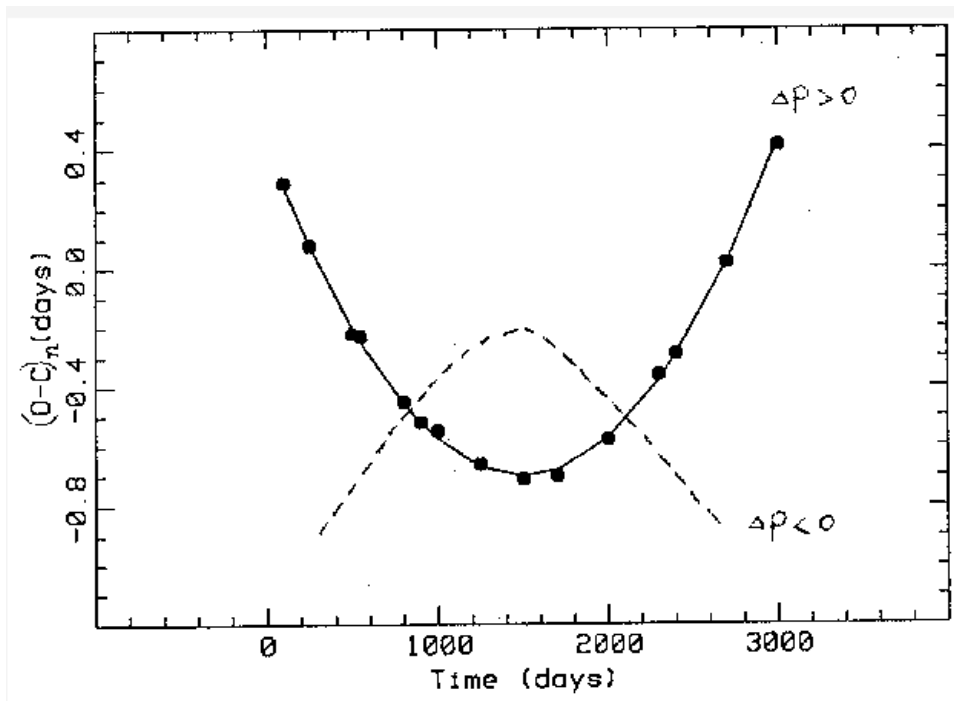,width=133mm,height=100mm,angle=-0.0}
  \parbox{80mm}{{\vspace{2mm} }}
  \end{center}\vspace{-10mm}
  \caption{The O--C diagram as the period is changing slowly with time.
$(O-C)_{n}=\triangle P\,n^{2}$. The solid curve for $\triangle P>0$,
the dashed curve for $\triangle P<0$. }
\label{fig:OC2}
\end{figure}

\noindent
Now, if the period is changing with time, may be increasing or decreasing.
Let P(t) stands for a changing period with a slow, constant rate $\beta$ as
caused by stellar evolution, we write
\begin{equation}
  P(t)=P_{0} + \beta t
\label{eq:pt1}
\end{equation}
The derivative of the equation above gives the rate of period changing
$ \frac{dP(t)}{dt}=\beta $.
The stellar radius or mass changes due to evolutionary effects or mass loss
often results in a slowly changing in period. We have
\begin{equation}
\begin{array}{l}
  C_{n}=T_0+nP_{0}, ~~~O_{n}=T_0+nP(T_{n})\\
  (O-C)_{n}=n[P(T_{n})-P_{0}]= n \beta T_{n}, ~~~n=0, 1, 2, \cdots\cdots, N
\end{array}
\end{equation}
By denoting $\triangle$P as the mean change during one period,
when $T_{1}=P, T_{2}=2P, \cdots\cdots, T_{N}=NP$, we obtain
\begin{equation}
  P(T_{1})=P_{0}+\triangle P, \cdots\cdots, P(T_{N})=P_{0}+N\triangle P
\end{equation}
So during the time span of the observations concerned, the change of the period
is
\begin{equation}
  (O-C)_{n}=n[P(T_{n})-P_{0}]=\triangle P\,n^{2}
\label{eq:OCn6}
\end{equation}
This is a parabola equation (see Fig.~\ref{fig:OC2}).
%
%

With assumption that
\begin{equation}
  C_{n_{0}}=T_0+P_{0}\,n_{0}, ~~~O_{n_{0}}=T_0(n_{0})+ P_{0}\,n_{0}
\end{equation}
we have
\begin{equation}
  (O-C)_{n}=(O-C)_{n-n_{0}}+(O-C)_{n_{0}}=\triangle P(n-n_{0})^{2}+(T_0(n_{0})-T_0)
\end{equation}
Now the equation demonstrates the information of both the epoch $T_0$ and the
period change $\triangle$P are included in the O--C diagram. The measured
accuracy of period changes increases as $n^{2}$. Hence it is more important
to extend the observations for a well-studied variable star.

The predicted evolutionary period changes of variable stars is commonly
difficult to be revealed in the O--C diagram because the evolutionary changes
may still be too small to be detected after a few tens of years. In that cases,
the O--C diagram should be essentially a straight line. In special, in those
stars evolving to the red toward the end of their horizontal branch lifetimes,
the evolutionary period changes, while still small, ought to be large enough
to be detected. In these cases, the O--C diagram should be well represented by
a parabolic curve, which should indicate a slow rate of period increase.
A few stars might be caught in the interval of core instability, and could
show period increases or decreases much larger than $dP/dt$=0.025 days per
million years.

Prager (1939) noted that three field RR Lyrae stars' periods did not change
at a constant rate, but were better modeled assuming abrupt period changes.
Abruptness, of course, does not accord with a model in which period change
results from smooth evolution. Usually, the period changes revealed in the
O--C diagram were more abrupt than a perfect parabola would allow.
The period changes are small compared to the length of the period itself but
because many cycles have passed by, the cumulative effects on the O--C diagram
may be large. In some stars, the observations can be well fit by assuming
that the period is changing at a constant rate (parabola O--C diagram), but
many others follow Prager's stars seem to have abrupt changes in period.
When fit by a parabola, the derived rates of period change are usually in the
range of $-$1 to +1 days per million years, or a change of less than 10 seconds
during 50--100 years in which the stars have been under observation.

\subsection{The Phase-shift Diagram}
\noindent
In addition to the O--C diagram, an analogy, the phase-shift diagram is also
applied to interpret period changes in light (e.g. Percy et al. 1980,
Breger 1990). Followings show the phase-shift diagram functions exactly
the same as the classical O--C diagram. We define quantity $f_{i}$ as below:
\begin{equation}
  f_{i}\equiv \frac{t_{i}-t_{0}}{P_{\rm est}}
\end{equation}
here $t_{i}$ is the observed times such as times of maximum light of pulsating
stars, $t_{0}$ is any arbitrary reference time, $P_{\rm est}$ is an estimated
period. The fraction remained of $f_{i}$ constructs the phase $\phi_{i}$,
and the integer of $f_{i}$ produces the cycle number N. A plot of the phase
against the cycle number is thus called the phase-shift diagram. Under the
assumption that the period P is changing uniformly at a rate of $\beta$ described
by eq.(\ref{eq:pt1}) where $P_{0}$ is the period value at $t_{0}$, t is the time
interval passed by from $t_{0}$, hence
\begin{equation}
  P_{i}=P_{0} + \beta (t_{i}-t_{0}), ~~~~i=1, 2, 3,\cdots\cdots, N
\end{equation}
If we take the mean of $P_{0}$ and $P_{i}$, the period at middle point of the
two periods is
\begin{equation}
  \langle P \rangle = \frac{1}{2}(P_{i}+P_{0})=P_{0}+\frac{1}{2}\beta(t_{i}-t_{0})
\label{eq:Pmean3}
\end{equation}
we have cycle numbers from $P_{0}$ to $P_{i}$:
$$ n=\frac{t_{i}-t_{0}}{\langle P \rangle}=\frac{t_{i}-t_{0}} { P_{0}+\frac{1}{2}\beta(t_{i}-t_{0}) },~~~~
   (t_{i}-t_{0})=\frac{n\,P_{0}}{1-\frac{1}{2}\beta n} $$
Such that we find $\phi_{n}$:
\begin{equation}
  \phi_{n}=\frac{t_{i}-t_{0}}{P_{\rm est}} - n = \frac{n\,P_{0}}{P_{\rm est}(1-\frac{1}{2}\beta n)} - n
\end{equation}
Without loss of generality, $\frac{1}{2}\beta n \ll 1$, we make use of series
expansion that
\begin{equation}
  \frac{1}{1-\frac{1}{2}\beta n} = \sum_{n=0}^\infty (\frac{1}{2}\beta n)^{n}
  \approx 1+\frac{1}{2}\beta n
\end{equation}
therefore,
\begin{equation}
  \phi_{n}=\frac{nP_{0}}{P_{\rm est}}(1+\frac{1}{2}\beta n)-n
      =(P_{0}-P_{\rm est})\frac{n}{P_{est}} + \frac{1}{2P_{\rm est}}P_{0}\beta n^{2}
      =(P_{0}-P_{\rm est})\frac{n}{P_{\rm est}} + \frac{1}{2}\beta n^{2}
\label{eq:percy51}
\end{equation}
($P_{0}/P_{\rm est}\approx 1, ~n=1, 2, 3,\cdots\cdots, N $). The equation presents
a parabola about cycle number n.
Quantity $\beta$, is generally used to characterize the period change rates
of variables. It is the rate of period change determined through fitting
a parabola to the O--C diagram, and is normally presented in the unit of
days/days or days/year. A unit of days per million years for $\beta$ is
chosen by some authors. $\frac{1}{P}\frac{dP}{dt} $ is sometimes presented
as the period change rate in unit cycle/year.

\section{Analytic O--C Curves}
\noindent
We describe the light brightness of a variable star with a sinusoidal waveform as:
\begin{equation}
  L(t)=L_{0}+A\sin(2\pi t/P)
\end{equation}
where P is the period of light variation, A is the amplitude of the
sinusoidal wave, $L_{0}$ is the zero point. It is well known that the times
of maximum light occur at $L'(t)=0$, which yields $O_{n}=\{ T_{0}, T_{1},
T_{2}, \cdots\cdots, T_{N} \}$. At the same time,
$ \sin\frac{2\pi C_{n}}{P} \longrightarrow$ 1 and $\cos\frac{2\pi C_{n}}{P}
\longrightarrow$ 0, at the times of maximum light.  The period changes in
the light curves are discussed in several different sources below.

\subsection{Frequency Modulation}
\noindent
For frequency or period modulation, theoretical light curve is defined by
\begin{equation}
  L_{\rm f}(t)=L_{0}+A\sin(2\pi t/P(t))
\end{equation}
$L_{\rm f}'(t)$=0 produces $O_{n}$ as
\begin{equation}
  O_{n}=\frac{P(t)}{4} (2n+1),~~n=0, 1, 2, \cdots\cdots, N
\end{equation}
$C_{n}$ are naturally derived from
$L(t)=L_{0}+A\sin(2\pi t/P)$, so
\begin{equation}
  (O-C)_{n}=\frac{1}{4} (2n+1)(P(t)-P),~~~~~~n=0, 1, 2, \cdots\cdots, N
\end{equation}
Recalling eq.(\ref{eq:pt1}): $P(t)=P+\beta t=P+\beta Pn$ which means $\beta$ is
the period change over one cycle, thus
\begin{equation}
  (O-C)_{n}=\frac{1}{2}\beta P n^{2} + \frac{1}{4} \beta P n
\end{equation}
this is a parabola with respect to cycle number n.

\subsection{Amplitude Modulation}
\noindent
The change in period caused by the variability of amplitude called amplitude
modulation is described as:
\begin{equation}
  L_{\rm a}(t)=L_{0}+A(t)\sin(2\pi t/P), ~~~A(t)\equiv a \sin(2\pi t/P_{\rm A})
\end{equation}
A(t) refers to a sinusoidal amplitude modulation which causes a secondary
period $P_{\rm A}$ in the light curve. The calculus of $L_{\rm a}$(t), $L_{\rm a}'(t)$,
gives observed times of maximum light $O_{n}$:
\begin{equation}
  \frac{1}{P_{\rm A}}\cos\frac{2\pi O_n}{P_{\rm A}}\sin\frac{2\pi O_n}{P} +
  \frac{1}{P  }\sin\frac{2\pi O_n}{P_{\rm A}}\cos\frac{2\pi O_n}{P}   =0
\end{equation}
By making use of the replacement of $O_{n}=(O-C)_{n} + C_{n}$ just for the
triangle functions involving P and expanding them.
Without loss of generality we presume
\begin{equation}
  (O-C)_{n}\ll P;~~~~  \sin\frac{2\pi (O-C)_{n}}{P} \approx \frac{2\pi (O-C)_{n}}{P}
\end{equation}
and with the help of the features mentioned in the beginning of this section:
$  \sin\frac{2\pi C_{n}}{P} \longrightarrow 1 $, we find
$$\frac{1}{P_{\rm A}} \cos\frac{2\pi O_{n}}{P_{\rm A}}
  ( \frac{2\pi (O-C)_{n}}{P} \cos\frac{2\pi C_{n}}{P} +
    \cos\frac{2\pi (O-C)_{n}}{P}
  ) +                        $$
\begin{equation}
  \frac{1}{P    } \sin\frac{2\pi O_{n}}{P_{\rm A}}
  ( \cos\frac{2\pi (O-C)_{n}}{P} \cos\frac{2\pi C_{n}}{P} -
    \frac{2\pi (O-C)_{n}}{P} ) =0
\end{equation}
Expanding similarly now the two functions including $P_{\rm A}$,
and recalling that $\cos\frac{2\pi C_{n}}{P} \longrightarrow$ 0,
$\cos\frac{2\pi (O-C)_n}{P} \longrightarrow$ 1, $L_{\rm a}'(t)=0 $ becomes
$$
 \frac{1}{P_{\rm A}} \cos\frac{2\pi (O-C)_{n}}{P_{\rm A}} \cos\frac{2\pi C_{n}}{P_{\rm A}} -
 \frac{1}{P_{\rm A}} \frac{2\pi (O-C)_{n}}{P_{\rm A}} - $$  $$
 \frac{1}{P    } \frac{[2\pi (O-C)_{n}]^{2}}{PP_{\rm A}}
   \cos\frac{2\pi C_{n}}{P_{\rm A}} -
 \frac{1}{P    } \frac{2\pi (O-C)_{n}}{P} \cos\frac{2\pi (O-C)_{n}}{P_{\rm A}} =0   $$
Note that the coefficient of the third term of the equation above is
infinitely small at higher order of $\frac{(O-C)_{n}}{P}$ which is set to
almost zero as our original consideration. This yields
\begin{equation}
 2\pi(O-C)_{n} (- \frac{1}{P_{\rm A}^{2}} - \frac{1}{P^{2}}
   \cos\frac{2\pi (O-C)_{n}}{P_{\rm A}} ) +
 \frac{1}{P_{\rm A}} \cos\frac{2\pi (O-C)_{n}}{P_{\rm A}} \cos\frac{2\pi C_{n}}{P_{\rm A}} =0
\label{eq:274}
\end{equation}
When $P_{\rm A}$ is so large enough that $\frac{(O-C)_{n}}{P_{\rm A}} \ll$ 1, we can
make following simplifications:
\begin{equation}
 \cos\frac{2\pi (O-C)_{n}}{P_{\rm A}} \approx 1- \frac{2\pi (O-C)_{n}}{P_{\rm A}},~~~
 \frac{[2\pi (O-C)_{n}]^{2}}{P_{\rm A}} \approx 0
\end{equation}
Consequently, we get
\begin{equation}
  (O-C)_{n}=\frac{ P_{\rm A}}{2\pi} \frac{ \cos\frac{2\pi C_{n}}{P_{\rm A}} }
  { 1 + (\frac{P_{\rm A}}{P})^{2} + \cos\frac{2\pi C_{n}}{P_{\rm A}} }
\label{eq:amp_mod}
\end{equation}
This relation shows the shape of O--C is predominated by the modulation period
$P_{\rm A}$. If $P_{\rm A}\approx P$, like cycle-to-cycle modulation,
$$ (O-C)_{n} \approx \frac{P_{\rm A}}{2\pi} $$
that indicates a shift of $P_{\rm A}/2\pi$ with respect to the zero line showing
constant periodic light. In the case of great difference between $P_{\rm A}$ and
P, if $P_{\rm A}\gg P$, the long-term amplitude modulation, stimulation shows
only a tiny scatter around the zero of the O--C, while $P_{\rm A}\ll P$,
the high frequency amplitude modulation, the feature of the O--C is of the
behaviour of periodic amplitude variation, we have after neglecting the term including P
in eq.(\ref{eq:274}),
\begin{equation}
  (O-C)_{n}= \frac{P_{\rm A}}{2\pi} \cos\frac{2\pi (O-C)_{n}}{P_{\rm A}} \cos\frac{2\pi C_{n}}{P_{\rm A}}
\label{eq:276a}
\end{equation}
A sinusoidal light curve with an amplitude modulation may have a significant
variation in O--C, while a saw-tooth light curve with the same amplitude
modulation will show an absolutely flat O--C.

\subsection{Double Mode Case}
\noindent
As for double-mode variables with small amplitudes, we choose:
$$ L_{\rm d}(t)=L_{0}+Asin\frac{2\pi t}{P_{1}}+Bsin( \frac{2\pi t}{P_{2}}+\phi )$$
Let $L_{\rm d}'(t)$=0 for finding $O_{n}$ and so $(O-C)_{n}$.
Replacing $O_{n}$ with $(O_{n}-C_{n})+C_{n}=(O-C)_{n}+C_{n}$.
With assumption of $(O-C)_{n}\ll P_{1}$, so
$\cos\frac{2\pi (O-C)_{n}}{P_{1}} \longrightarrow 1 $,
$\sin\frac{2\pi (O-C)_{n}}{P_{1}} \approx \frac{2\pi (O-C)_{n}}{P_{1}} $ and
$\sin\frac{2\pi (O-C)_{n}}{P_{2}} \approx \frac{2\pi (O-C)_{n}}{P_{2}} $,
and recall that we want to find $P_{2}$ in the O--C diagram.
Therefore, the expanding of $L_{d}'(t)$=0 is equivalent to
$$
  \frac{P_{2}}{P_{1}}\cdot\frac{A}{B}\cdot\frac{2\pi (O-C)_{n}}{P_{1}}
  \sin\frac{2\pi C_{n}}{P_{1}} +
  \frac{2\pi (O-C)_{n}}{P_{2}} \sin( \frac{2\pi C_{n}}{P_{2}} + \phi )
= \frac{P_{2}}{P_{1}}\cdot\frac{A}{B}\cdot \cos\frac{2\pi C_{n}}{P_{1}} +
  \cos( \frac{2\pi C_{n}}{P_{2}} + \phi )  $$
thus we have $(O-C)_{n}$ like
\begin{equation}
  (O-C)_{n}=\frac{1}{2\pi}
\frac{ \frac{P_{2}}{P_{1}}\frac{A}{B} \cos\frac{2\pi C_{n}}{P_{1}} +
       \cos( \frac{2\pi C_{n}}{P_{2}} + \phi )
     }
     { \frac{P_{2}}{P_{1}^{2}} \frac{A}{B} \sin\frac{2\pi C_{n}}{P_{1}} +
       \frac{1}{P_{2}} \sin( \frac{2\pi C_{n}}{P_{2}} + \phi )
     }
\end{equation}
At the times of maximum light, we know
$    \cos\frac{2\pi C_{n}}{P_{1}} \longrightarrow 0,~
    \sin\frac{2\pi C_{n}}{P_{1}} \longrightarrow 1.  $~
Finally, we get
\begin{equation}
  (O-C)_{n}=\frac{P_{2}}{2\pi}
\frac{ \cos( \frac{2\pi C_{n}}{P_{2}} + \phi )
     }
     { (\frac{P_{2}}{P_{1}})^{2}\frac{A}{B} +
       \sin( \frac{2\pi C_{n}}{P_{2}} + \phi )
     }
\label{eq:ocdm}
\end{equation}
For the two different periods, $P_{2}/P_{1}\sim 0.78$ is the normal case
for the large-amplitude $\delta$ Sct stars. Taking a common amplitude ratio
value of 0.30 and a second period of 0.03 d we may simplify eq.(\ref{eq:ocdm})
to be
$$  (O-C)_{n} \sim \frac{0.005\cos( \frac{2\pi C_{n}}{P_{2}} + \phi )  }
                        {0.2+ \sin( \frac{2\pi C_{n}}{P_{2}} + \phi )  }$$
Beating phenomenon can also be involved in the feature of the O--C curve when
the two periods are closely spaced with near amplitudes. Furthermore,
the effects of the beating will be particularly obvious in the times of
maximum light, giving a large O--C amplitude. An interference of two
oscillations with the same amplitude A for a suspect $\beta$ Cep star,
53 Psc (Wolf 1987), leads to the amplitude modulation of the form:
\begin{equation}
  A_{\rm b}=A\sqrt{ 2(1+cos\frac{2\pi t}{P_{\rm b}}) }, ~~~
  \frac{1}{P_{\rm b}}=\frac{1}{P_{2}} - \frac{1}{P_{1}}
\label{eq:Pb}
\end{equation}
here $P_{b}$ is the beat period. Based on this equation, the time required
for the amplitude to reach a certain value $A_{obs}$, counted from the epoch
of $A_{\rm b}$=0, can be derived as
\begin{equation}
  \triangle t \approx \frac{P_{\rm b} A_{\rm b}}{2\pi A}
\end{equation}
One more example is the light-time effect in a binary system if
$P_{2} \gg P_{1}$, from which the O--C curve may be interpreted,
\begin{equation}
  (O-C)_{n}=\frac{P_{1}^{2}}{2\pi P_2}\cdot\frac{B}{A}
  \cos( \frac{2\pi C_{n}}{P_{2}} + \phi )
\label{eq:286a}
\end{equation}

The information revealed by the O--C diagram depends on both `O' constructed
from maxima (often used for pulsating stars) or minima (as in binary) and
the methods used to derive the `O' from the data. Usually four ways to find `O',
we take
\begin{itemize}
  \item the time of maximum brightness for each single observation;
  \item the time of maximum brightness in a smoothed light curve;
  \item the phase of the maximum of a representative mean light curve fitted
to all or a portion of the observations over that cycle;
  \item the time of maximum derived from an exact solution for a Fourier series which
has been fitted to the data.
\end{itemize}

\section{The Light--Time Effect of A Binary}
\noindent
If a pulsating star is a member of a binary system, an independent estimate
of the mass from stellar pulsation theory can be given. This is the intention
of the detection of the light-time effect (LTE) of a pulsating star.

\subsection{Examples of Three $\beta$  Cep Stars}
\noindent
Pigulski has throughly investigated the nature of the period changes of three
large-amplitude $\beta$ Cephei-type stars: $\beta$ Cephei itself (the prototype
of this class), $\sigma$ Scorpii and BW Vulpeculae in terms of the combined
O--C diagram constructed from both the light and radial velocity data. It was
shown that the observed period changes were caused by the superposition of a
constant-rate period change originated from normal stellar evolution and of
the light-time effect in a binary system.

Pigulski and Boratyn (1992) proposed that the observed changes of $\beta$
Cephei (=HR 8238=HD 205021, V=3.23, B1 IV) can be understood in terms of
orbital motion of the variable in a binary system, in which the secondary
is the less massive companion resolved by speckle interferometry. In fact,
the companion in the system of $\beta$ Cep was found in the speckle
interferometric way by Gezari et al. (1972). By assuming that the orbital
motion is the unique cause of the observed variation of the pulsation period,
\begin{equation}
  P_{\rm puls}=P_{0}(1+\frac{V_{\rm rad}}{c})
\end{equation}
$V_{\rm rad}$ is fully described by six elements of the spectroscopic orbit:
$\gamma$, radial velocity of the system's barycentre; $P_{\rm orb}$, orbital
period; $K_{1}$, semi-amplitude of the radial velocity curve, due to orbital
motion of the variable; $e$, eccentricity of the orbit; $\omega$, longitude
of periastron; $T_{0}$, the time of periastron passage.
The analyses of a single O--C diagram constructed in terms of both the
photometric and radial velocity times of maximum offered by Pigulski and
Boratyn (1992) show that the apparent changes of the pulsation period of
$\beta$ Cep can be explained by the light-time effect, induced by the
orbital motion of the variable around the barycentre of the variable-speckle
companion system.

$\sigma$ Scorpii (=HR 6084=HD 147165, B1 III) is a quadruple system. Its light
variability can be described as a superposition of four periodic terms(
Jerzykiewicz \& Sterken 1984), of which the main pulsation period $P_{2}$=
$0^{d}$.24684 and $P_{1}$=0$^{d}$.23967 are also observed in radial velocity.
The combined O--C diagram using the times of maximum radial velocity and
light displays both an increase and a decrease of the main pulsation period.
The period changes seen in the O--C curves can be explained neither by an
evolutionary effect nor by LTE alone.
\begin{equation}
  (O-C)_{\rm LT}=(O-C)+ 0^{d}.000015\,E- 1^{d}.29\times 10^{-10}\, E^{2}
\label{eq:OC_LT}
\end{equation}
In fact, both effects contributed to the observed changes of the main
pulsation period of $\sigma$ Sco. And LTE is caused by
the speckle tertiary. Due to the observations do not cover the whole orbital
period of the tertiary, the $(O-C)_{\rm LT}$ diagram cannot help to give a more
accurate value of the orbital period than that provided by Evans et al.~(1986),
is of the order of 100--350 yr. In this sense, the spectroscopic elements
are unavailable from the $(O-C)_{\rm LT}$ diagram. But it is sure that the
changes seen in the $(O-C)_{\rm LT}$ diagram can be fully explained by the
orbital motion (Pigulski 1992).

\begin{figure}
  \vspace{-28mm}
  \begin{center}
  \hspace{2mm}\psfig{figure=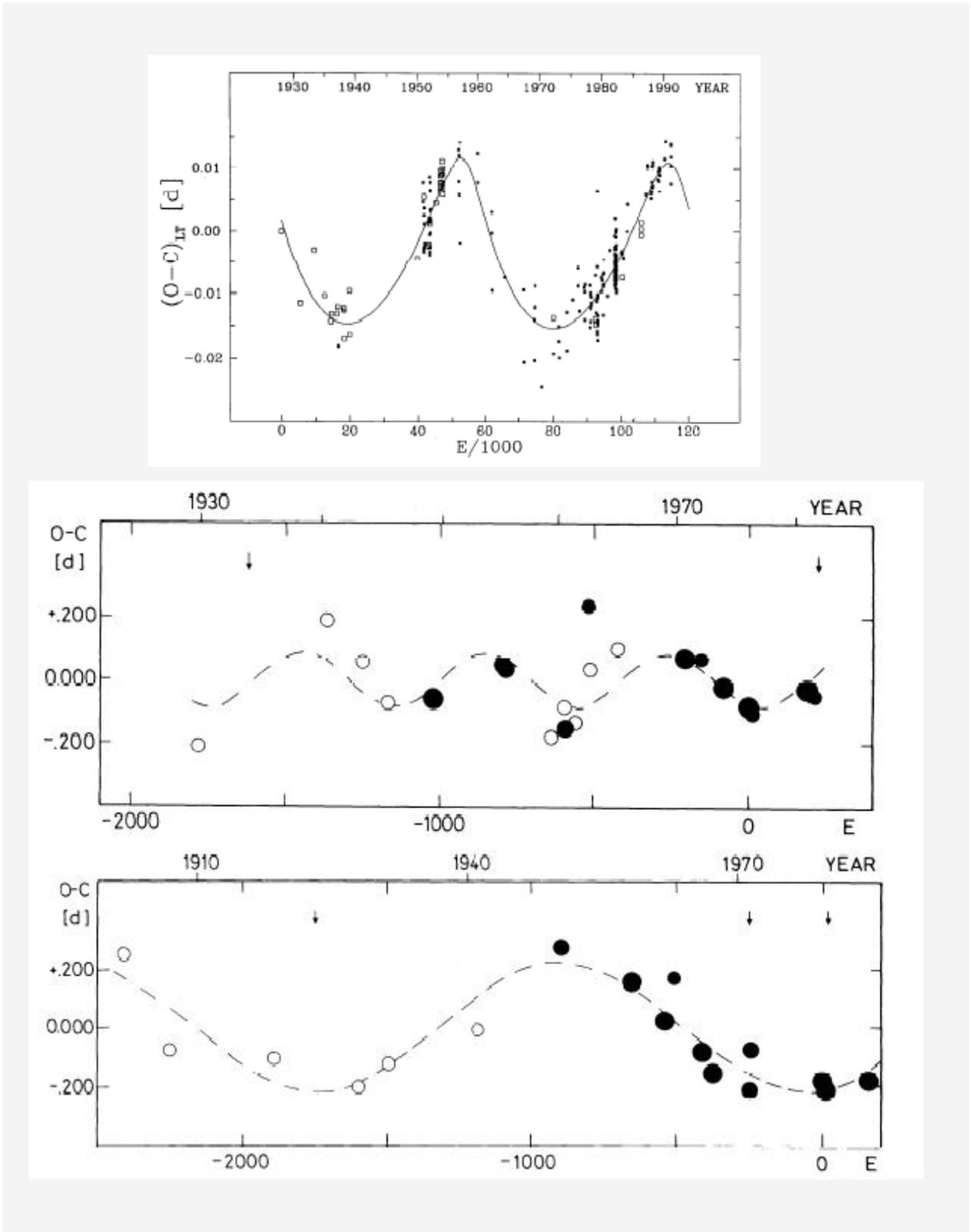,width=160mm,height=210mm,angle=-0.0}
  \end{center}
  \caption{The O--C diagrams showing the light-time effect.
Upper panel: $(O-C)_{\rm LT}$ for the $\beta$ Cep star BW Vul, the changes seen in
this figure are interpreted as the light-time effect only. Continuous line
represents the best fit, from which spectroscopic elements of the orbit are
determined. After Pigulski (1993). Lower panels: sinusoidal variations in
the O--C curves of the Cepheids FN Aql and RX Aur. Open circles denote
the residuals obtained from photographic observations; photoelectric
measurements are marked with dots. The size of these symbols refers to the
weight assigned. Courtesy of Szabados (1988).}
\label{fig:OC_LTBW}
\end{figure}

BW Vulpeculae (=HR 8007=HD , V=6.55, B2 III) is a monoperiodic large-amplitude
$\beta$ Cep star. Odell (1984) and Jiang Shi-Yang (1985) showed that
the parabolic fit to the O--C diagram of BW Vul yields periodic residuals.
Both authors attributed these residuals to LTE induced by
the motion of the star in a binary system with an orbital period of about
25 yr. Pigulski (1993) showed that the combined effect of the increase of
period with a rate of +2.34 s/cen and LTE in a binary
system ($P_{\rm orb}$=33.5$\pm$0.4 yr) fully accounts for the observed changes
of the pulsation period of BW Vul. The light-time effect in the $(O-C)_{\rm LT}$
diagram is, for the time being, the only observational evidence for the
presence of a companion to BW Vul (see Fig.~\ref{fig:OC_LTBW}).
For BW Vul, the times of minimum light and of minimum radial velocity were
used in the determination of period changes. One ephemeris of minimum radial
velocity was adopted so as to produce a single O--C diagram.
%
%

All periodic variables with stable periods such as pulsating stars, eclipsing
binaries and pulsars offer the possibility of detecting unseen companion(s)
by means of the light-time effect.
A pulsating star is a promising candidate for the detection of LTE provided
that it belongs to a binary or multiple system. A simplest case is the LTE in
a double system with a circular orbit. In the O--C diagram, the changes of
the period P of a signal emitted by the visible component (pulsating
primary) due to LTE will have the form of a sine curve with semi-amplitude
equal to $P_{\rm orb}\,K_{1}/2\pi c$, where $2K_{1}$ is the range of the primary's
radial velocity curve, $P_{\rm orb}$ is the orbital period. The amplitude of LTE
depends proportionally on $P_{\rm orb}$ and $K_{1}$ and is equal to the time
which light needs for passing the projected orbit. For a given orbital period,
short-period variables are more suitable for the study of LTE than the
long-period ones because the visibility of LTE depends on the $P_{\rm orb}/P$
ratio. Pigulski (1995) listed 10 objects including eclipsing binaries for
which period changes attributable to LTE are observed.

\subsection{The Binary Model of The $\delta$ Sct Star CY Aqr}
\begin{figure}
 \vspace{-40mm}
  \begin{center}
  \hspace{-10mm}\psfig{figure=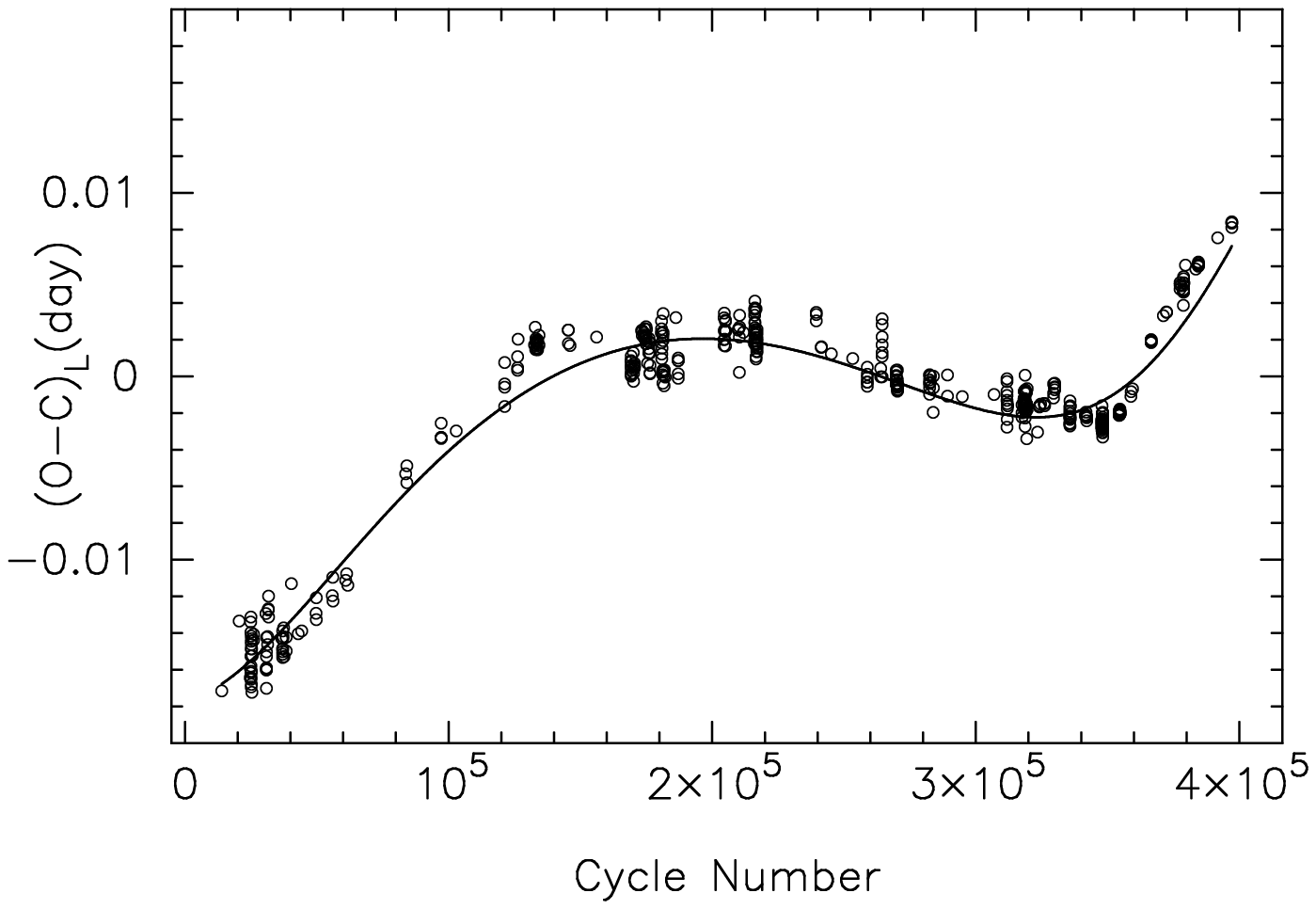,width=150mm,height=110mm,angle=-90.0}
  \end{center}
  \vspace{-25mm}
  \caption{The O-C diagram of CY Aqr (cycles) and the curve fitted using
both a parabolic curve and a trigonometric function (solid line). }
\label{fig:CYAqr3f}
\end{figure}

\begin{figure}
  \vspace{-40mm}
  \begin{center}
  \hspace{-10mm}\psfig{figure=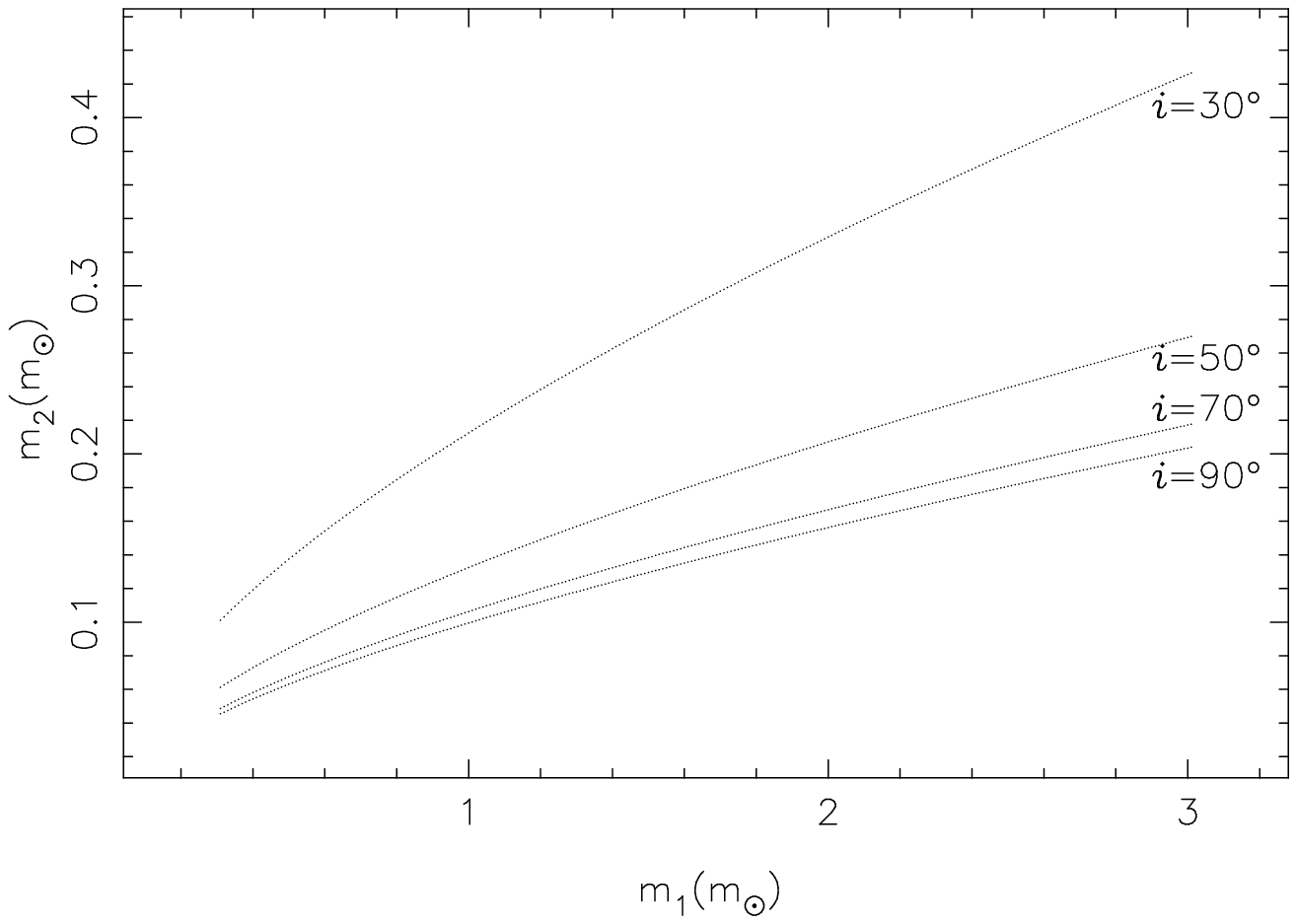,width=150mm,height=110mm,angle=-90.0}
  \end{center}
  \vspace{-30mm}
  \caption{Solutions to the mass function for various choices
of $\sin i$ . }
\label{fig:CYAqr5f}
\end{figure}

\noindent
The possibility that CY Aqr is a spectroscopic binary (see Fig.~\ref{fig:CYAqr3f})
immediately leads to the hope that a mass can be determined.
In Figure~\ref{fig:CYAqr5f} we show the relationship between the masses of
the two companions for assumed values of the orbital inclination.
The curve for $i=90^{\circ}$ sets a lower limit to $m_{2}$ at each $m_{1}$.
Recent results given by McNamara et al.(1996) indicate:
$m_{1}$=1.06\,M$_{\odot}$, $M_{v}$=2.5. For this mass value of the primary star
$m_{2}$ rounds  0.15\,M$_{\odot}$. The companion must be fainter than CY Aqr by
about 8.8 mag. We stress that, the binary model seems to us to be the most
plausible interpretation of the variable times of maximum light. It remains
only a reasonable hypothesis until new confirming data are available. The
simplest and most easily accomplished test is to examine the radial velocity
of CY Aqr with a 1.4 km s$^{-1}$ amplitude predicted by the model.
%
%
%
%

\section{Period Change Noise}
\begin{figure}
  \vspace{-20mm}
  \begin{center}
  \hspace{-5mm}\psfig{figure=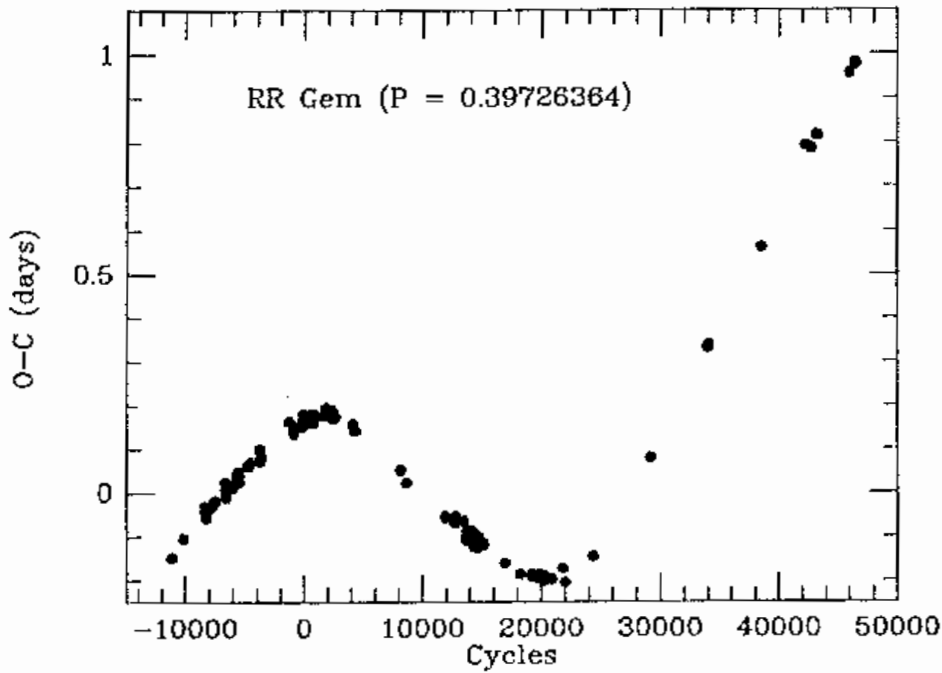,width=133mm,height=100mm,angle=-0.5}
  \end{center}
  \vspace{-10mm}
  \caption{The O--C diagram for the times of maximum light of RR Gem, after
Smith (1995).}
\label{fig:RRGem}
\end{figure}

\noindent
Some stars appear to be unchanged in period over the time span of observations;
others have undergone multiple changes in period. Some stars have exhibited
both an increase and a decrease in period, as can be seen in the O--C diagram
of RR Gem in Fig.~\ref{fig:RRGem}. Stellar evolution is not expected as we
know to act so quickly as to cause period changes in both directions in so
short a span. Moreover, the sizes of the observed period changes are often
much larger than expected from theory. We call this difference as
``period change noise''. In this regard, it is not unquestionable to attribute
the period changes observed in any individual star to evolution alone.

%
%

The period change noise is a result of random mixing events associated with
the semiconvective zone of the stellar core (Sweigart and Renzini 1979).
The changes in the internal structure of a variable star arise in two ways:
through the gradual change in core composition caused by nuclear burning,
but also through a composition redistribution in the deep interior caused by
convective overshooting and the formation of a semiconvective zone. Whereas
this composition redistribution is assumed in standard calculations to
proceed smoothly on the same timescales as nuclear burning. The nuclear
reaction and convection need not be perfectly coupled at every instant of
time. For an actual star, the composition redistribution is a discontinuous
process. There might then be many mixing events, each slightly altering the
internal of the star, and each leading to a change in period. Furthermore,
different types of mixing events can account for both period increasing and
decreasing. The observed timescales of period changes are in at least rough
agreement with the mixing event hypothesis. Nuclear burning would be expected
to produce a slow rate of period change which would be essentially ~unvarying
over the span of a century. The mixing events, on the other hand, if they
took place quickly would produce period changes which would seem abrupt; to
the contrary, on a longer timescale they might produce period changes which,
for one hundred years, would seem constant in rate. Of course, over a
sufficiently long period of time, the period changes due to nuclear burning
and discrete mixing events at the semiconvective zone must average to the
value predicted by those evolutionary calculations which assume that the
composition redistribution takes place ~smoothly.

As another ~origin for the period change noise, proposed by Stothers (1980),
which invoking hydromagnetic effects (such as for RR Lyrae stars). It is a
common phenomenon involving stellar magnetism.

Third, Dearborn et al.~(1990) suggested that exotic particles, such as weakly
interacting massive particles (WIMPS), might produce changes in the pulsation
periods (say of RR Lyrae stars). The existence of these particles has been
postulated to explain the apparent existence of large amounts of dark matter
in galaxies. These particles might become trapped within stars, where they
can provide a new mechanism of energy transfer. In the case of horizontal
branch stars, WIMPS might cause thermal pulses to occur on the
Kelvin-Helmholtz timescale for the stellar core. These thermal pulses
would produce period changes. The timescale calculated for these changes,
however, on the order of $10^{6}$ yr, is too long to explain the shorter
term period noise.

Fourth, the light travel time effect because of orbital motion in a binary
stars system may be another possible source for
apparent period changes as in RR Lyrae variables identified by Coutts (1971),
and in $\beta$ Cephei stars inspected by Pigulski \& Boratyn (1992) and Pigulski (1992, 1993).
The motion of a star in a binary system could introduce a long term periodic
oscillation in the O--C diagram for the time of maximum light, as the star
alternately approaches and ~recedes from us. It is difficult, however, to
build a case for the binary nature of any ~variables ~from O--C data alone.
The examples are comparatively few. Saha and White (1990) have argued
on the basis of O--C data and radial velocity observations that
TU UMa, a RR Lyrae stars, may be a binary; Barnes III and Moffett(1975) and
Moffett et al.~(1987, 1988) evidenced the duplicity of SZ Lyncis,
a dwarf Cepheid; others as BS Aqr (Fu Jiang-Ning et al. 1997),
CY Aqr(Zhou Ai-Ying \& Fu Jiang-Ning 1998), BE Lyn (Kiss \& Szatmary 1995,
Liu et al. 1994, Tang et al. 1992, Liu et al. 1991), BW Vul (Pigulski 1993)
and so on (we refer the reader to examples preceded by Pigulski 1993, 1995).
However, does the binary hypothesis by itself provide an adequate explanation
for the period change behavior?

Finally, mass loss might alter the theoretical period changes. For stars on
the horizontal branch lost their masses when they were in the instability strip,
that is, when they were RR Lyrae stars. But, the predicted rates of period
changes were not significantly changed for mass loss rates as high as
$10^{-9}$M$_{\odot}$/yr.

\section{Period Change to Metallicity}
\noindent
As a variable star evolves through the H--R diagram, crossing its various
stages of evolution, the metal abundance or chemical composition changes
as well as its period does. In other words, at each given effective
temperature periods vary with metallicity. Thus it is natural there is some
correlation between the two observable quantities. In real, the period change
depends on the change of chemical abundance. We write the relationship of
the period change with respect to metallicity, P---[Fe/H], as
$$   \frac{dP}{dt} =\frac{dP}{dt}([Fe/H], P) $$

The spectral type determined from the hydrogen Balmer lines (termed Sp(H)) was
often later than the spectral type determined from the prominent Ca II K line
at 3933\AA (termed Sp(K)), which investigated by Preston (1959) under a low
resolution spectroscopic survey of RR Lyrae stars in the solar vicinity.
On the basis of the difference, Preston then proposed $\triangle S$ index as
an indicator of metal abundance (hereafter the $\triangle$S method):
\begin{equation}
    \triangle S= 10 [Sp(H) - Sp(K)]
\end{equation}
which was later on confirmed by the analyses of high resolution spectra of
RR Lyrae stars (Preston 1961; Butler 1975) and of model stellar spectra (Manduca 1981).
Butler(1975) derived
\begin{equation}
    [Fe/H]= -0.23 - 0.16\triangle S
\end{equation}
Blanco(1992) ~further improved above as
\begin{equation}
    [Fe/H]= -(0.34\pm0.02) - (0.18\pm0.05) \triangle S
\end{equation}
Based on the globular cluster metallicity scale (adopted by Zinn \& West
1984, and Zinn 1985), Suntzeff et al.~(1991) got following calibration,
\begin{equation}
    [Fe/H] = -0.408 - 0.158 \triangle S
\end{equation}
is about 0.2 dex more metal-poor than Butler's scale.
In the case of RRab stars, lower $\triangle$S (high metallicity), smaller
period changes; higher $\triangle$S (low metallicity), bigger period changes.
Iron abundance [Fe/H] is ~inferred from the strength of the calcium H-line.
Certainly, such a calibration does not require that [Fe/H] = [Ca/H].
Thus, it does require that there exist a one-to-one correlation between iron
and ~calcium abundances. In fact, [Ca/H] appears to be slightly high compared
to [Fe/H] for metal-poor stars in general. Manduca (1981) provided an
approximate relation as
\begin{equation}
    [Ca/H] = 0.8 [Fe/H]
\end{equation}
for theoretical calibration of the $\triangle$S index.

In addition to the method, other metallicity indicators have also been
developed. The photometric indices among which are Sturch's(1966) broadband
ultraviolet blanketing index $\delta$(U-B), the photometric k-line index
$(k-b)_{2}$ (Jones 1971, 1973), the Str\"{o}mgren $m_{1}$ index (Epstein 1969,
Epstein and Epstein 1973), and the Walraven system $\delta$[B-L] index (Lub
1977, 1979). In brief, metallicity can be estimated from $\triangle S$(Preston),
$(k-b)_{2} $(Jones), $\triangle (B-L)$ (Walraven), $\delta (U-B)$ (Sturch) and
$\delta m_{1}$ (Str\"{o}mgren photometry). For the range
0$\leq \triangle$ S$\leq$12, we have relationships below among the approaches:
\begin{equation}
  \begin{array}{l}
    \triangle S=-84\delta(U-B) + 14  \\
    \triangle S=-44(k-b)_{2} + 10\\
    \triangle S=-99(m_{1})_{0} + 16
  \end{array}
\end{equation}
(Butler 1975) and
\begin{equation}
  \triangle S=11.9 -194\triangle [B-L]+815(\triangle [B-L])^{2}
\end{equation}
(Blanco 1992).
If metallicity did not change over the pulsation cycle of the star, these
indices would remain constant. Rodr\'{\i}guez et al.(1990) failed a test of
this internal consistency in the fact of SX Phe and the large amplitude
$\delta$ Sct stars. Instead, they concluded that the $m_{1}$ index variation
is larger when the metal abundance is smaller.

\section{Ending Remarks}
\noindent
The O--C diagram provides us the accurate period and maximum epoch of a
variable star, a detector of its period and phase changes, and the information
about the spectroscopic orbital elements if it is a binary.
Plus, the O--C data provide constraints on stellar evolution theory.
For binary systems, O--C analyses of eclipses can reveal very gradual
changes due to tidal interactions, or even to gravitational radiation.

The information of period changes is also presented as a phase diagram
instead of a classical O--C diagram.
Various shapes of the O--C curves correspond to following different cases:
\begin{enumerate}
   \item The O--C diagram for a star without measurable changes in period is
a straight line.
   \item If the period of the star is constant, and if the correct period has
been adopted, points on the O--C diagram will scatter around a straight
horizontal line (see Fig.~\ref{fig:OC1}).
   \item If the period of the star is constant, but the adopted period is too
long or too short, the straight line will slop up or down (see Fig.~\ref{fig:OC1}).
   \item If the period of the star is changing at a slow, constant rate (i.e.
$P(t)=P_{0}+\beta t$ where $\beta$ is small), then a good approximation of
the O--C diagram can be represented by a parabola (see Fig.~\ref{fig:OC2}).
   \item If the period change is caused by the light-time effect in a binary
system, the light curve appears to be sinusoidal (see Fig.~\ref{fig:OC_LTBW}).
   \item Broken lines---abrupt or jump changes of period (see Fig.~\ref{fig:RRGem}).
\end{enumerate}

In order to interpret an O--C diagram correctly, it is necessary to know a
number of cycles have elapsed between two observed maxima or since an
initial epoch. The determination of elapsed cycles from an epoch is not
always easy and reliable especially while the period of the star has been
changing significantly.

All common period-finding techniques assume that the periods are constant
over the time span of the observations and that the phases are stable.
However, these conditions are not always met by variables. Therefore,
the O--C method is, in fact, one of the necessities of period analysis.
A short review on the O--C method we refer to Willson (1986), a detailed
theoretical study may be also referred to the paper series published in
MNRAS (Koen \&Lombard 1993, 1995; Lombard \& Koen 1993; Koen 1996).
Koen and Lombard had paid scrupulous attention to the O--C methodology.
They pointed out that the problem suffered from within the O--C method that
the series of intervals among times is non-stationary so that the applications
of this technique may lead to spurious conclusions. As well, they modelled
the period changes in variable stars on a quantitative basis, especially
supplied tests for the presence of intrinsic scatter in the periods of
the long-period pulsating stars.

\smallskip
\large
\noindent
\bf
Acknowledgements\\
\rm
\normalsize
\noindent
This work was supported by the National Natural Science Foundation of China.

\smallskip
\large
\bf
\noindent References\\
\rm

\footnotesize
\vspace{-2.1mm}
\noindent
Barnes III T. G. and Moffett T. J., 1975, AJ, 80, 48\\
Blanco V., 1992, AJ, 104, 734\\
Breger M., 1990, in Confrontation Between Stellar Pulsation and
  Evolution, eds. C. Cacciari and G. Clementini, ASP Conf. Ser., Vol.11, p.263\\
Butler D., 1975, ApJ, 200, 68\\
Coutts C. M., 1971, in New Directions and New Frontiers in Variable Star
  Research, veroff. der Remeis-Sternwarte Bamberg, IX, Nr. 100, 238\\
Dearborn D., Raffelt G., Salat P., Silk J., and Bouquet A., 1990, ApJ, 354, 568\\
Epstein I., 1969, AJ, 74, 1131\\
Epstein I. and Epstein A. E. A., 1973, AJ, 78, 83\\
Evans D. S., Mcwilliam A., Sandmann W. H., Frueh M., 1986, AJ, 92, 1210\\
Fu Jiang-Ning, Jiang Shi-Yang, Gu Sheng-Hong, Qiu Yu-Lei, 1997, IBVS No.4518\\
Gezari D. Y., Labeyrie A., Stachnik R. V., 1972, ApJ, 173, L1\\
Jerzykiewicz M. and Sterken C., 1984, MNRAS, 211, 297\\
Jiang Shi-Yang, 1985, Acta Astrophys. Sinica, 5, 192\\
Jones D. H. P., 1971, MNRAS, 154, 79\\
Jones D. H. P., 1973, ApJS, 25, 487\\
Kiss L.I. and Szatmary K., 1995, IBVS No.4166\\
Koen C., 1996, MNRAS, 283, 471(IV)\\
Koen C. and Lombard F., 1993, MNRAS, 263, 287(I)\\
Koen C. and Lombard F., 1995, MNRAS, 274, 821(III)\\
Liu Yan-Ying, Jiang Shi-Yang, Cao Ming, 1991, IBVS No.3606\\
Liu Zong-Li and Jiang Shi-Yang, 1994, IBVS No. 4077\\
Lombard F. and Koen C., 1993, MNRAS, 263, 309(II)\\
Lub J., 1977, A\&AS, 29, 345\\
Lub J., 1979, AJ, 84, 383\\
Manduca A., 1981, ApJ, 245, 248\\
McNamara, D. H., Powell, J. M., and Joner, M. D. 1996, PASP, 108, 1098\\
Moffett T.J., Barnes III T.G., Fekel F.C., et al., 1987,
  Bull. Am. Astron. Soc., 19, 1085\\
Moffett T.J., Barnes III T.G., Fekel F.C., et al., 1988, AJ, 95, 1534\\
Odell A. P., 1984, PASP, 96, 657\\
Percy, J. R., Mattews, J. M., and Wade, J. D. 1980, A\&A, 82, 172\\
Pigulski A., 1992, A\&A, 261, 203  \\
Pigulski A., 1993, A\&A, 274, 269\\
Pigulski A. and Boratyn D. A., 1992, A\&A, 253, 178\\
Pigulski A., 1995, in Astronomical and Astrophysical Objectives of
  Sub-Milliarcsecond Optical Astronomy. eds. E. H$\o$g and P. K. Seidelmann,
  IAU Symp., Vol.166, p.205\\
Prager R., 1939, Harv. Bull., No.911, 1\\
Preston G. W., 1959, ApJ, 130, 507\\
Preston G. W., 1961, ApJ, 134, 633\\
Rodr\'{\i}guez E., L\'{o}pez de Coca P., Rolland A., Garrido R.,
  1990 , Rev. Mexican Astron Astrof., 20, 37\\
Saha A. and White R. E., 1990, PASP, 102, 148(erratum: 102, 495)\\
Smith H. A., 1995, RR Lyrae Stars, Cambridge University Press, Cambridge\\
Stothers R., 1980, PASP, 92, 475\\
Sturch C., 1966, ApJ, 143, 774 \\
Suntzeff N. B., Kinman T. D., and Kraft R. P., 1991, ApJ, 367, 528\\
Sweigart A. V. and Renzini A., 1979, A\&A, 71, 66\\
Szabados L., 1988, PASP, 100, 589\\
Tang Qing-Quan, Yang Da-Wei, Jiang Shi-Yang, 1992, IBVS No.3771  \\
Willson L. A., 1986, in The Study of Variable Stars using Small
  Telescopes, ed. John R. Percy, Cambridge University Press, Cambridge, p.219\\
Wolf M., 1987, IBVS No.3003\\
Zhou Ai-Ying and Fu Jiang-Ning, 1998, Delta Scuti Star Newsletter, 12, 28(Vienna)\\
Zinn R., 1985, ApJ, 293, 424\\
Zinn R. and West M. J., 1984, ApJS, 55, 45\\

\end{document}